\documentclass[prc,twocolumn,showpacs,nofootinbib]{revtex4-1}
\usepackage{amsmath,amssymb,amsfonts}
\usepackage{graphicx}
\usepackage{hhline}
\usepackage{bm}
\usepackage{amsmath}
\usepackage{epsfig}

\usepackage[dvips]{color}

\begin{document}         
\title{Nuclear response functions with finite-range Gogny force: tensor terms
  and instabilities} 
\author {A. De Pace}
\affiliation{Istituto Nazionale di Fisica Nucleare, Sezione di Torino, 
Via P. Giuria 1, I-10125 Torino, Italy}
\author {M. Martini}
\affiliation{ESNT, CEA, IRFU, Service de Physique Nucl\'eaire, Universit\'e de Paris-Saclay, F-91191 Gif-sur-Yvette Cedex, France} 

\begin{abstract}
A fully-antisymmetrized random phase approximation calculation employing the
continued fraction technique is performed to study nuclear matter response
functions with the finite range Gogny force.  
The most commonly used parameter sets of this force, as well as some recent
generalizations that include the 
tensor terms are considered and the corresponding response functions are shown. 
The calculations are performed at first and second order in the continued
fraction expansion and the explicit expressions for the second order  
tensor contributions are given. Comparisons between first and second order
continued fraction expansion results are provided.  
The differences between the responses obtained at the two orders turn out
to be  
more pronounced for the forces including tensor terms than for the standard
Gogny ones.  
In the vector channels the responses calculated with Gogny forces including
tensor terms are characterized by a large heterogeneity,  
reflecting the different choices for the tensor part of the interaction. 
For the sake of comparison the response functions obtained considering a $G$-matrix
based nuclear interaction are also shown. 
As a first application of the present calculation, the possible existence of
spurious finite-size instabilities of the Gogny forces with or without tensor
terms has been investigated. The positive conclusion is that all the Gogny
forces, but the GT2 one, are free of spurious finite-size instabilities. 
In perspective, the tool developed in the present paper can be inserted in the
fitting procedure to construct new Gogny-type forces.  
\end{abstract}
\pacs{21.30.Fe, 21.60.Jz, 21.65.-f}
\maketitle 
\section{Introduction}

The past years have been characterized by a regain of interest in nuclear
matter response functions in connection with the nuclear structure. 
The power of this tool to study electron, neutrino, meson and nucleon
scattering on nuclei in the quasielastic peak and beyond 
(see for example
Refs.~\cite{Alberico:1981sz,Krew88,DeP93,Gil97,DePace:1996fb,Ichi06,Martini:2009uj})
is well known.
Recently the formalism of the response functions has been largely employed to
investigate the properties of effective nuclear forces. 
After the seminal work of Ref.~\cite{GarciaRecio:1991cp}, a lot of efforts have
been done in connection with the Skyrme forces in order to
generalize the fully-antisymmetrized Random Phase Approximation (RPA)  
results obtained in that paper, since those calculations were limited to
consider the central and the density-dependent terms of the Skyrme interaction.  
The first generalization has been the inclusion in the RPA calculations of the
spin-orbit term \cite{Margueron:2006wh} and after that several papers have been
devoted to the inclusion of the tensor term \cite{Davesne:2009ky}  
and to the investigation of its role \cite{Pastore:2012mf,Pastore:2012dq}, 
up to the generalization of the formalism to asymmetric nuclear matter
\cite{Davesne:2014yaa}. The effects of other density-dependent terms
\cite{Pastore:2014yua} 
and of the three-body forces \cite{Davesne:2014vua} have also been treated.
A review on this topic recently appeared \cite{Pastore:2014aia}. 
The two main applications of this kind of calculations have been, up to now, 
the study of the unphysical finite-size instabilities in nuclear energy density
functionals
\cite{Lesinski:2006cu,Pastore:2012mf,Pastore:2012dq,Hellemans:2013bza,Pastore:2014yua,Pastore:2015vfa} 
and the calculation of the neutrino mean free path in nuclear and neutron matter
\cite{Iwamoto:1982zp,Reddy:1998hb,Navarro:1999jn,Margueron:2006wh,Pastore:2012dq,Pastore:2014yua}.

At variance with the case of the Skyrme forces, less attention has been paid
to the case of finite range Gogny forces. The reason is probably twofold.
First, in spite of the successes of this kind of force (with which pairing
correlations can be automatically taken into account in the mean-field based
calculation, without the introduction of further parameters), the number of
mean-field based calculations using the Gogny force is enormously inferior to
the number of the corresponding calculations with zero-range Skyrme forces.  
The second reason is that, due to the finite range of the force, fully
analytical calculations of the antisymmetrized RPA nuclear matter response are no
longer possible because of the role played by the exchange terms.
With finite range forces analytical results can be achieved only in the so
called ring approximation, which takes into account only the direct
contributions, or by considering the Landau-Migdal (LM) limit.  
The first paper devoted to the RPA response function in infinite nuclear matter
employing the Gogny force has been the one of Gogny and Padjen
\cite{GognyNPA77}, which followed the LM approach. 
An approximation beyond the standard LM one, based on keeping the full momentum 
dependence in the direct term and making the LM approximation for the exchange 
term (LAFET), has been studied for the Gogny force in
Ref.~\cite{Margueron:2005nc} and compared with the results of a method,
developed in the same paper, based on an expansion of the Bethe-Salpeter
equation onto a spherical harmonic basis, a method that in principle can be
carried out up to any degree of accuracy.  

Another approach for the treatment of the fully antisymmetrized RPA response
with finite range nuclear forces is the one based on the continued fraction (CF)
technique. 
Calculations employing this method, for meson-exchange type potentials, have
been done in
Refs.~\cite{Alberico:1993ur,Barbaro:1993jg,Barbaro:1995ez,Barbaro:1995gp} by 
truncating the CF expansion at the first order.
The calculation has then been pushed up to the second order in
Ref.~\cite{DePace:1998yx}.  
The CF technique up to the second order has been employed also in
Ref.~\cite{Margueron:2008dq}, using a Gogny force.   
Their results support the ones of Ref.~\cite{DePace:1998yx} on the rapid 
convergence of the CF expansion founding, among other, that up to the
saturation density the convergence is already achieved at first order. 

In the present paper we use the CF technique, following the approach of 
Ref.~\cite{DePace:1998yx}, and we make use of the Gogny force.  
Our aim is the study of the nuclear matter response function by employing the
most commonly used parametrizations of the Gogny force (D1 \cite{Gogny:1973},
D1S \cite{Berger:1991zza}, D1N \cite{Chappert:2008zz}, D1M
\cite{Goriely:2009zz}), as well as the recent generalizations that include the
tensor terms
\cite{Otsuka:2006zz,Anguiano:2011kz,Anguiano:2012mx,DeDonno:2014roa}. 
Several recent papers have shown the crucial role played by the tensor term in
the behavior of the nuclear matter response functions 
\cite{Davesne:2009ky,Pastore:2012mf,Pastore:2012dq,Davesne:2014yaa} and, as a
consequence, on the finite-size spurious instabilities, but all these studies
have only considered Skyrme energy density functionals. Very few works cope with
the role of the tensor in the Gogny force. To our knowledge, the only two papers
related to this subject are Ref.~\cite{Pastore:2013sja}, which performs RPA 
calculations of the response functions, but in the LM approximation, showing 
results for the D1MT force, and Ref.~\cite{Navarro:2013bda}, which considers, for
the D1ST and D1MT interactions, the spin susceptibilities, \textit{i.e.} the $q
\to 0$, $\omega \to 0$ limit of the nuclear responses.

\section{Formalism}

In the evaluation of response functions we shall employ Green's functions
techniques, as described e.~g. in Refs.~\cite{Fet71,Abr63}.
We consider an infinite system of nucleons at a density corresponding to a
Fermi momentum $k_F$, interacting through a non-relativistic potential, whose
general form, in momentum space, reads
\begin{eqnarray}
  V(\bm{k}) &=& V_0(k) +
    V_{\tau}(k) \bm{\tau}_1\cdot\bm{\tau}_2 +
    V_{\sigma}(k) \bm{\sigma}_1\cdot\bm{\sigma}_2 \nonumber \\
  && \quad  + V_{\sigma\tau}(k) \bm{\sigma}_1\cdot\bm{\sigma}_2 \ 
      \bm{\tau}_1\cdot\bm{\tau}_2 +
    V_t(k) S_{12}(\hat{\bm{k}}) \nonumber \\ 
  && \quad +
    V_{t\tau}(k) S_{12}(\hat{\bm{k}}) \bm{\tau}_1\cdot\bm{\tau}_2 ,
\label{eq:pot}
\end{eqnarray}
where $S_{12}$ is the standard tensor operator and $V_{\alpha}(k)$ represents
the momentum space potential in channel $\alpha$ (here we neglect, for sake of
simplicity the spin-orbit terms).

The response of the system to an external probe can be obtained from the
particle-hole (ph) four-point Green's function $G^{\text{ph}}$, which is the
outcome of a Galitskii-Migdal integral equation:
\begin{widetext}
\begin{eqnarray}
  && G^{\text{ph}}(K+Q,K;P+Q,P) = 
    -G(P+Q)\,G(P)\,(2\pi)^4\delta(K-P) \nonumber \\
  && + i G(K+Q)\,G(K)\int\frac{d^{4}T}{(2\pi)^4}
    \Gamma^{\text{ph}}(K+Q,K;T+Q,T)\,
    G^{\text{ph}}(T+Q,T;P+Q,P) , \nonumber \\
\label{eq:Gph}
\end{eqnarray}
\end{widetext}
$G$ being the exact one-body Green's function and $\Gamma^{\text{ph}}$ the
irreducible vertex function in the ph channel (capital letters here refer to
four-vectors; small case letters to three-vectors). For brevity, in
Eq.~(\ref{eq:Gph}) we have dropped the spin-isospin indices.

Out of $G^{\text{ph}}$ one can define the polarization propagator in a given
spin-isospin channel $X\equiv(S,M,T)$:
\begin{eqnarray}
  \Pi_{X}(Q) & \equiv & 
    \Pi_{X}(q,\omega) \\
  &=&  i\int\frac{d^{4}P}{(2\pi)^4}\frac{d^{4}K}{(2\pi)^4}
    G^{\text{ph}}_{X}(K+Q,K;P+Q,P) \nonumber .
\label{eq:Pi}
\end{eqnarray}
Finally, the system response functions are simply proportional to the imaginary
part of $\Pi_{X}$:
\begin{eqnarray}
  R_{X}(q,\omega) 
  &=& -\frac{V}{\pi} \text{Im}\Pi_{X}(q,\omega) \nonumber \\
  &=& -\frac{3\pi A}{2k_F^3} \text{Im}\Pi_{X}(q,\omega) ,
\label{eq:RX}
\end{eqnarray}
$V$ being the volume and $A$ the mass number of the system.

Depending on the approximations done on $G$ and $\Gamma^{\text{ph}}$, one can
get different approximations for $\Pi_{X}$.
By neglecting $\Gamma^{\text{ph}}$ and dressing the nucleon propagators with
the first order self-energy $\Sigma^{(1)}(k)$ one obtains the Hartree-Fock (HF)
or mean field approximation, $\Pi^{\text{HF}}$. 
Here we shall follow the usual approximation of including the mean field
effects through the HF effective mass
\begin{eqnarray}
  \frac{m^*}{m} = \left(1+\frac{m}{k_F}
    \frac{\partial\Sigma^{(1)}(k)}{\partial k}\Bigg|_{k=k_F}
    \right)^{-1} ,
\end{eqnarray}
$m$ being the bare nucleon mass.
By replacing in Eq.~(\ref{eq:Gph}) the irreducible vertex function
$\Gamma^{\text{ph}}$ with the matrix elements of the bare potential, one gets
the RPA equation for $G^{\text{ph}}$.
However, one still has a closed equation only for the full four-point Green's
function and not for the simpler polarization propagator.

In order to get the RPA response functions we shall follow the approach based
on the CF expansion of the polarization propagator
\cite{Len80,Fes92,Del85,Del87,DePace:1998yx}. An alternative approach, based on
the CF expansion of the effective interaction, has been developed in
Refs.~\cite{Sch89,Margueron:2008dq}.
Up to second order in the CF expansion (the highest order so far reached in
actual calculations) the two approaches are equivalent.
Details of the derivation can be found in Ref.~\cite{DePace:1998yx}. Here we
summarize the relevant formulae.

\begin{figure}
\begin{center}
\includegraphics[clip,width=0.48\textwidth]{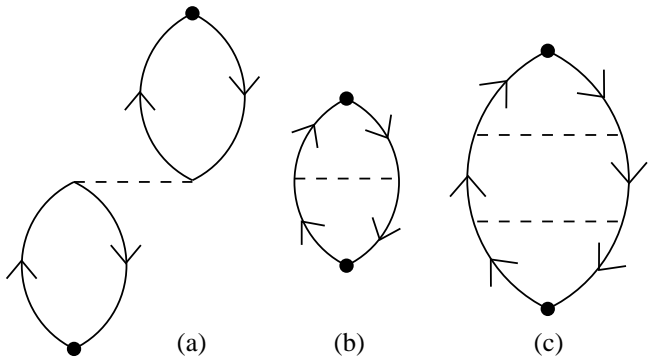}
\caption{Feynman diagrams that enter the second order CF expansion of the RPA
  propagator: first-order direct (a), first order exchange (b) and second order
  exchange (c).}
\label{fig:RPA} 
\end{center}
\end{figure}

In the CF expansion the RPA polarization propagator reads
\begin{widetext}
\begin{equation}
  \Pi^{\text{RPA}}_{\text{X}} = \frac{\Pi^{\text{HF}}}
    {\displaystyle 1 
    -\Pi^{(1)\text{d}}_{\text{X}}\big/\Pi^{\text{HF}}
    -\Pi^{(1)\text{ex}}_{\text{X}}\big/\Pi^{\text{HF}}
    -\frac{\strut 
      \Pi^{(2)\text{ex}}_{\text{X}}\big/\Pi^{\text{HF}}
      -\left[\Pi^{(1)\text{ex}}_{\text{X}}\big/\Pi^{\text{HF}}
      \right]^2}{\displaystyle 1-...}} ,
\end{equation}
\end{widetext}
where the expansion has been explicitly shown up to second order.
$\Pi^{(1)\text{d}}_{\text{X}}$, $\Pi^{(1)\text{ex}}_{\text{X}}$ and
$\Pi^{(2)\text{ex}}_{\text{X}}$ correspond to the Feynman diagrams a, b and c
of Fig.~\ref{fig:RPA}, respectively.

The $n$-th order exchange propagator in the spin-isospin channel X reads 
\begin{equation}
  \Pi^{(n)\text{ex}}_{\text{X}}(q,\omega) = 
    \sum_{\alpha_i}
    C^{\alpha_1}_{\text{X}} \dots C^{\alpha_n}_{\text{X}} 
    \Pi^{(n)\text{ex}}_{\alpha_1...\alpha_n}(q,\omega) ,
\end{equation}
where the indices $\alpha_i$ run over all the spin-isospin channel of the
interaction and all the spin-isospin factors depending on the probed channel
are condensed in the coefficients $C^{\alpha_i}_{\text{X}}$ (see
Table~\ref{tab:Calphas}). 
Thus, the calculation of the RPA response at order $n$ in the CF expansion is
reduced to the calculation of the exchange contributions
$\Pi^{(n)\text{ex}}_{\alpha_1...\alpha_n}$ up to order $n$.
Details about the latter in the case of a Gogny interaction are given in the
Appendix. Analogous expressions for the case of a meson-exchange potential can
be found into the Appendix of Ref.~\cite{DePace:1998yx}. 

\begin{table}
\begin{center}
\begin{tabular}{ccccccc}
  \toprule
  $\strut\phantom{\Big|}X\equiv(S,M,T)$ & $C^0_X$ & $C^\tau_X$ & $C^\sigma_X$ & 
  $C^{\sigma\tau}_X$ & $C^t_X$ & $C^{t\tau}_X$ \\ \colrule
  $(0,0,0)$ & 1  &  3  &  3  &  9  &  0  &  0  \\ 
  $(0,0,1)$ & 1  & -1  &  3  & -3  &  0  &  0  \\ 
  $(1,1,0)$ & 1  &  3  & -1  & -3  & -1  & -3  \\ 
  $(1,1,1)$ & 1  & -1  & -1  &  1  & -1  &  1  \\ 
  $(1,0,0)$ & 1  &  3  & -1  & -3  &  2  &  6  \\ 
  $(1,0,1)$ & 1  & -1  & -1  &  1  &  2  & -2  \\ 
  \botrule
\end{tabular}
\end{center}
\caption{ The spin-isospin coefficients $C^{\alpha}_X$ (see text), in the 
various spin-isospin channels, for the interaction of
Eq.~(\protect\ref{eq:pot}). } 
\label{tab:Calphas}
\end{table}

\section{Response functions results}
\subsection{Standard parametrizations of the Gogny force}

The general expression of the Gogny interaction in the coordinate space is
\begin{equation}
{\begin{array}{lll}
V(\bm{r})&=&\displaystyle{\sum_{j=1}^{2}} 
  \left( W_j +B_j P_\sigma - H_j P_\tau - M_j P_\sigma P_\tau\right)
  e^{-\frac{r^2}{\mu_j^2}}\\    \\
&& +t_0 \left( 1+x_0 P_\sigma\right) \rho^{\alpha_0} \delta
  \left({\bm{r}}\right), 
\end{array}}
\label{Eq:Gogny_force}
\end{equation}
where $\bm{r}$ is the distance between two nucleons.
The first term of Eq. (\ref{Eq:Gogny_force}) is given by a sum of two Gaussians
with effective range $\mu_1$ and $\mu_2$ simulating the short- and long-range
components of a realistic interaction in the nuclear medium. This finite range
term includes all possible mixtures of spin and isospin operators, being  
$P_\sigma=(1+\bm{\sigma}_1\cdot\bm{\sigma}_2)/2$ and
$P_\tau=(1+\bm{\tau}_1\cdot\bm{\tau}_2)/2$ the spin and isospin exchange
operators, respectively. The second term is the zero-range density-dependent
contribution.
Usually the Gogny force also contains a zero-range spin-orbit term.
We omit to include it, as generally done in the nuclear matter calculations
\cite{Margueron:2005nc,Margueron:2008dq,Ventura:1994pnm,LopezVal:2006ef,Sellahewa:2014nia}. Indeed, in
Ref.~\cite{Margueron:2006wh} it was shown that the effects of the spin-orbit
interaction on the calculations of the nuclear response functions are small,
even at momentum transfer larger than the Fermi momentum.
This conclusion obtained by using Skyrme-type forces is expected to remain
valid also for the Gogny forces, due to the similarity of the spin-orbit term
in the two cases.
The expression of the Gogny interaction given in Eq. (\ref{Eq:Gogny_force}) is
the one corresponding to the most commonly used parametrizations, such as D1,
D1S, D1N and D1M. For the values of the $\mu_j, W_j, B_j, H_j, M_j, t_0, x_0,
\alpha_0$ parameters corresponding to these four forces see for example
Ref. \cite{Peru:2014yga}. 

\begin{figure}
\begin{center}
\includegraphics[clip,width=0.48\textwidth]{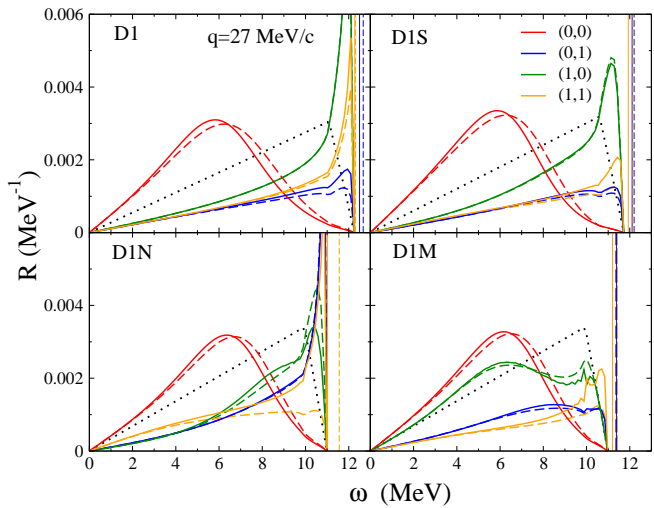}
\caption{(Color online) RPA response function in symmetric nuclear matter at
  $k_F=270\textrm{ MeV}/c$ and $q=27 \textrm{ MeV}/c$ calculated with the
  continued fraction technique at the first order (dashed line) and at the
  second order (solid line) in the CF expansion by considering different
  parametrizations of the Gogny interaction. The different spin and isospin
  channels $(S,T)$ are plotted in different colors. The black dotted lines
  represent the HF response. The collective modes above the continuum
    are represented by vertical lines. }
\label{fig:resp_q027_kf270_d1snm} 
\end{center}
\end{figure}

\begin{figure}
\begin{center}
\includegraphics[clip,width=0.48\textwidth]{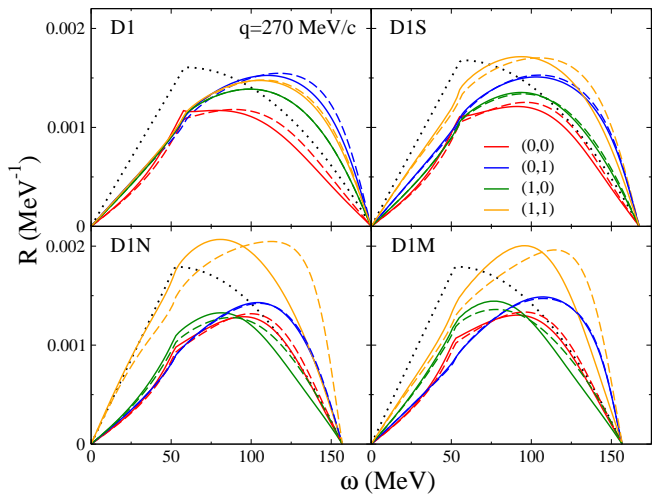}
\caption{(Color online) As in Fig.~\ref{fig:resp_q027_kf270_d1snm}, but for
  $q=270 \textrm{ MeV}/c$.}
\label{fig:resp_q270_kf270_d1snm} 
\end{center}
\end{figure}

Let us consider now the RPA response functions $R_{X}(q,\omega)$ in the four
spin and isospin channels calculated at first and second order in the CF
expansion using the four parametrizations of the Gogny interaction mentioned
above. For sake of illustration we show these responses in
Figs.~\ref{fig:resp_q027_kf270_d1snm} and \ref{fig:resp_q270_kf270_d1snm}
for the fixed value of $k_F$=270 $\textrm{MeV}/c$ and for two values of the
momentum transfer, $q=27$ $\textrm{MeV}/c$ and $q=270$ $\textrm{MeV}/c$. 
We consider these values in order to compare our results with the ones obtained
in Ref.~\cite{Margueron:2008dq} where only the D1 parametrization was
considered.  
Starting by this interaction one can observe that our results are similar to
the ones obtained in Ref.~\cite{Margueron:2008dq} (we remind that there is a
global multiplicative factor $4 \rho_0 \hbar c$ between our results and the
ones of Ref. \cite{Margueron:2008dq}). 
In analogy with Ref.~\cite{Margueron:2008dq} we can conclude that for the D1
parametrization the first order in the CF expansion gives a reliable
description of the  response functions for the $(S,T)$ spin and isospin
channels $(1,0)$ and $(1,1)$, but not for the $(0,0)$ channel, where it is
necessary to include the second order. For the $(0,1)$ channel, while at
$q=270$ $\textrm{MeV}/c$ the results of our work are once again similar to the
ones of Ref.~\cite{Margueron:2008dq} (and are characterized by a small effect
between the two orders of the CF expansion), some differences between the two
calculations appear at $q=27$ $\textrm{MeV}/c$. In our case, the collective
mode above the particle-hole continuum region remains outside the continuum
even at the second order in the CF expansion. 
It does not seem to be the case in the calculations of
Ref.~\cite{Margueron:2008dq}.

Turning to the other three interactions, whose results for the different
response functions are presented here for the first time, we can make two
general comments:
\begin{itemize}
  \item{} In the $(0,0)$ channel the responses calculated with the different
    parametrizations show a similar qualitative behavior, not only when
    compared to each other, but also with respect to the HF results; moreover, 
    the difference between the first and the second order CF expansion results is similar for the four forces.

  \item{} In the other $(S,T)$ channels, the various parametrizations can show
    important differences in the transferred-energy dependence; the convergence
    of the CF expansion as well strongly depends on the force parameters in the
    different $(S,T)$ channels.    
\end{itemize}

Considering now specific parametrizations, we can affirm, qualitatively
speaking, that the results obtained with D1S are not so different from the ones
obtained with D1.  
For these two parametrizations, some difference can be observed in the $(1,1)$
channel. Indeed, at $q=27$ $\textrm{MeV}/c$ the collective mode is above the p-h
continuum for D1S, but not for D1. At $q=270$ $\textrm{MeV}/c$ the response is
quenched with respect to the HF one in the case of D1, while there is a small
enhancement for the D1S case. In this $(1,1)$ channel, the difference between
the first and the second order in the CF expansion is more pronounced for D1S
than for D1.  

Remaining on this $(1,1)$ channel, the enhancement of the responses at $q=270$
$\textrm{MeV}/c$ with respect to the HF case, as well as the discrepancy between
the first and the second order in the CF expansion, are largely evident for D1M
and D1N. 

Without entering into the details of the behavior of the responses
in the different channels for each parametrization, we just mention that at
$q=27$ $\textrm{MeV}/c$ the response in the $(1,0)$ channel
calculated with D1M is quite different with respect to the other
parametrizations. 

In the $(0,1)$ channel the results strongly depend on the appearance 
(like in the case of D1, D1S and D1M) or
not (like in the case of D1N) of the collective mode above the ph continuum. 
This collective mode in the $S=0$, $T=1$ channel is the counterpart of the giant 
dipole resonance (GDR) in finite nuclei. 
A large scale theoretical QRPA calculation of dipole excitations in the whole
nuclear chart has been performed in Ref.~\cite{Martini:2016mqk} where the D1S
and D1M Gogny forces are used. 
A systematic shift toward lower energies is found for the GDR mode calculated
with D1M with respect to D1S. 
An equivalent shift appears in the present infinite nuclear matter results. 
Lower energy collective modes for D1M when compared to D1S are also found in
the $(1,1)$ channel, here for infinite nuclear matter, and in
Ref.~\cite{Martini:2014ura} for the Gamow-Teller resonances of finite nuclei. 
   
Beyond the discussed differences related to the use of different parameter sets
of the Gogny force, an important and general comment is in order:
independently of $q$, of the choice for the parametrization of the interaction 
and of the spin-isospin channel of excitation,
one can observe that at low $\omega$ the difference between the results
obtained at first order in the CF expansion and the ones obtained at second
order is always small. This is particularly important for the studies of the
finite size instabilities of the nuclear matter, that will be discussed in
Sect.~\ref{sec:instabilities}, which involve calculations of the response
functions at $\omega=0$.

\begin{widetext}

\begin{figure}[p]
\begin{center}
\includegraphics[clip,width=\textwidth]{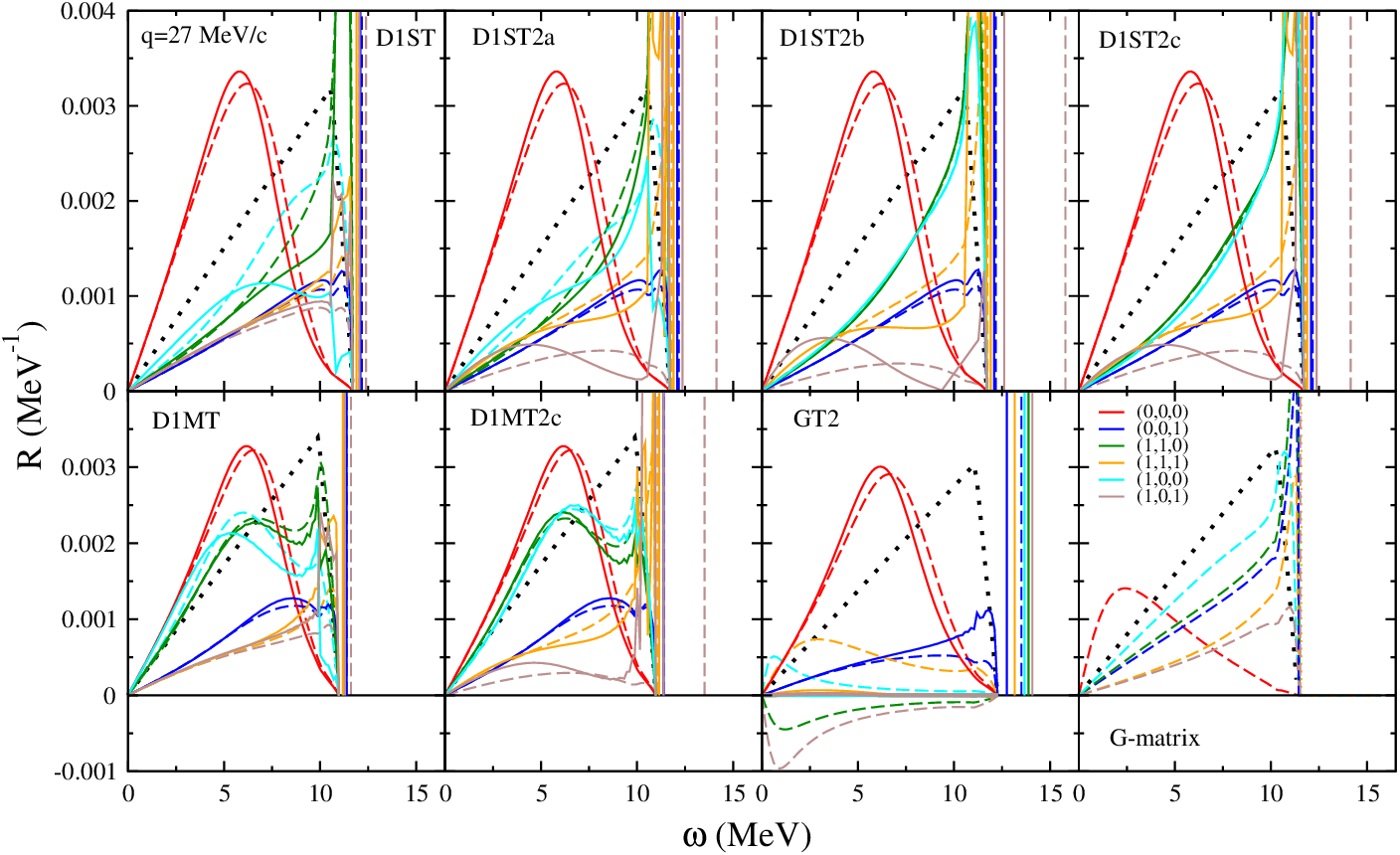}
\caption{(Color online) RPA response function in symmetric nuclear matter at
  $k_F=270$ $\textrm{MeV}/c$ and $q=27$ $\textrm{MeV}/c$ calculated with the
  continued fraction technique at the first order (dashed line) and at the
  second order (solid line) in the CF expansion by
  considering different parametrizations of the Gogny interaction with tensor
  terms. The different spin and isospin channels $(S,M,T)$ are plotted in
  different colors. The black dotted lines represent the HF response.
  The collective modes above the continuum
    are represented by vertical lines. For comparison, we also display the response functions calculated with a
  $G$-matrix nuclear interaction. }
\label{fig:resp_q027_kf270_gognytens} 
\end{center}

\begin{center}
\includegraphics[clip,width=\textwidth]{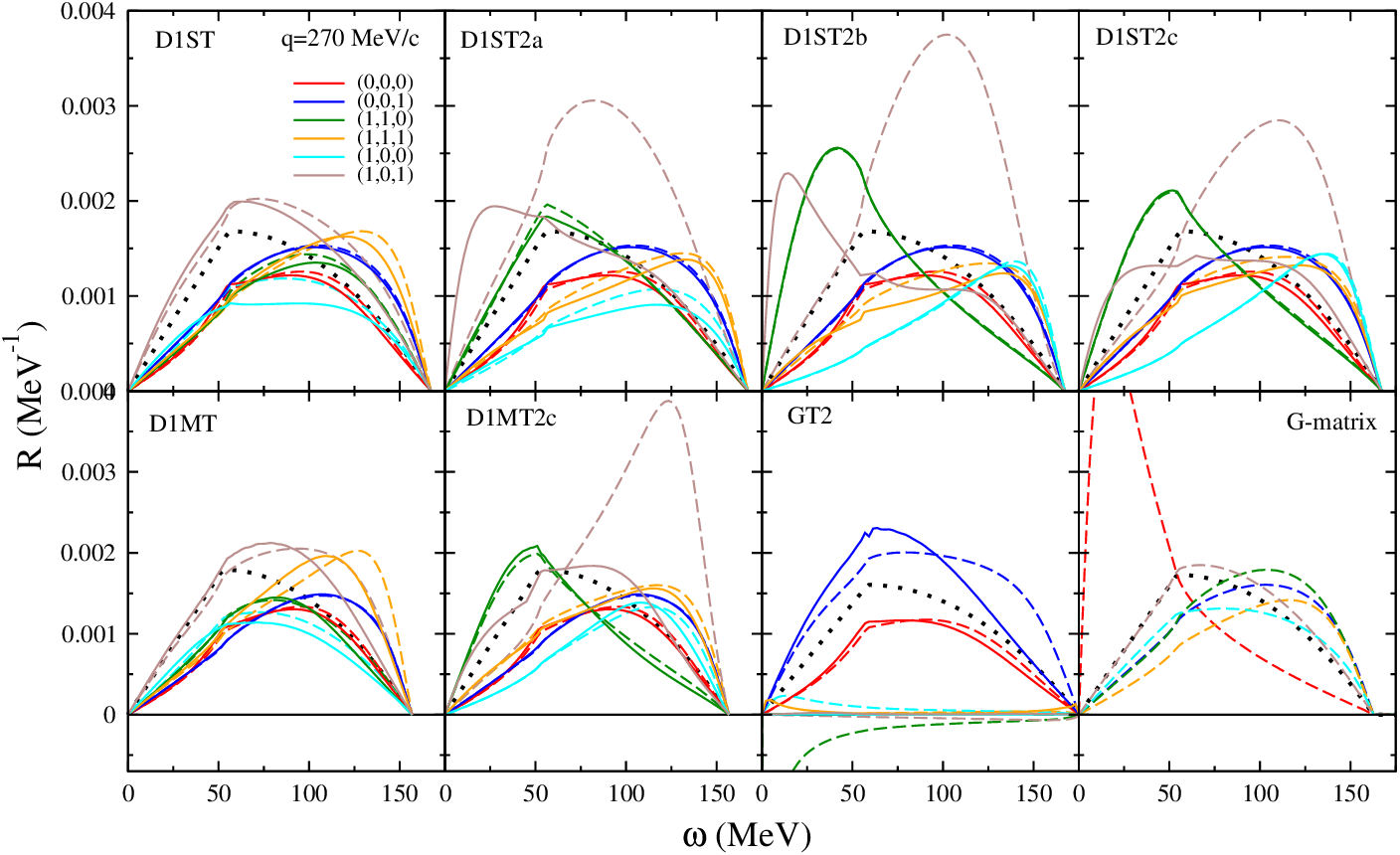}
\caption{(Color online) As in Fig.~\ref{fig:resp_q027_kf270_gognytens}, but for
  $q=270$ $\textrm{MeV}/c$.}
\label{fig:resp_q270_kf270_gognytens} 
\end{center}
\end{figure}

\end{widetext}

\subsection{Gogny force with tensor terms}

Some Gogny-type forces, less used up to now in the literature, are
characterized by the presence of additional tensor terms.
This is the case of the GT2 force \cite{Otsuka:2006zz} in which a
tensor-isovector contribution of Gaussian form 
is added to the central channels of Eq. (\ref{Eq:Gogny_force}). 
Tensor terms of Gaussian form appear also in the D1ST2a and D1ST2b
\cite{Anguiano:2012mx} as well as in the D1ST2c and D1MT2c
\cite{DeDonno:2014roa}, where beyond a tensor-isovector component a
tensor-isoscalar one is included.
Another important difference between the GT2 force and these
last four parametrizations is that in the GT2 case the inclusion of tensor
term involved a refitting of all the parameters whereas for the other cases the
tensor terms have been added to the D1S or D1M without changing the values of
the central parameters. In the same spirit also the D1ST and D1MT interaction
were introduced \cite{Anguiano:2011kz}. In this case the radial part of the
additional tensor-isospin term was based on the analogous one in the
microscopic Argonne V18 interaction, hence it was not characterized by a
Gaussian behaviour.  
In order to include also these D1ST and D1MT interactions in our analysis,
which is based on Gaussian interactions, we have fitted the tensor component of
the D1MT and D1ST interaction with a sum of three Gaussians.

Turning to the nuclear responses, we show in
Figs.~\ref{fig:resp_q027_kf270_gognytens} and \ref{fig:resp_q270_kf270_gognytens}
the results obtained at first and second order in the CF expansion by employing
the D1ST, D1ST2(a,b,c), D1MT, D1MT2c and GT2 Gogny interactions. 
For sake of comparison we also
display the response functions calculated at the first order 
employing a $G$-matrix based nuclear interaction. This interaction is based on
the $G$-matrix calculation of Ref.~\cite{Nak84} and it has been employed in RPA
calculations of quasielastic response functions both in finite nuclei
\cite{Shi89,Shi90,DePace:1996fb} and in nuclear matter \cite{DePace:1998yx}.
For $k_F$ and $q$ we choose the same values ($k_F$=270 $\textrm{MeV}/c$; $q=27$
$\textrm{MeV}/c$ and $q=270$ $\textrm{MeV}/c$) as the ones considered for
Figs.~\ref{fig:resp_q027_kf270_d1snm} and \ref{fig:resp_q270_kf270_d1snm}. 

The responses shown in Figs.~\ref{fig:resp_q027_kf270_gognytens} and
\ref{fig:resp_q270_kf270_gognytens} have never been calculated before by
considering the CF expansion approximation. 
However in the case of the D1MT interaction we can compare our results with the
ones published in Ref.~\cite{Pastore:2013sja}, where the responses for the D1MT
interaction have been calculated in the Landau framework by truncating the
residual interaction at $l_\textrm{max}=3$.  
Our results for $q\simeq 0.1 k_F$ are plotted in the left lower panel of
Fig.~\ref{fig:resp_q027_kf270_gognytens} and should be compared with the
results of Fig.~3 of Ref.~\cite{Pastore:2013sja}. 
The agreement is fairly good in all the $(S,M,T)$ channels. The agreement
between the two approaches no longer holds at $q\simeq k_F$, as one can notice
by comparing our results in the left lower panel of
Fig.~\ref{fig:resp_q270_kf270_gognytens} with the results of Fig.~4 of
Ref.~\cite{Pastore:2013sja}. 
The only channels where the two calculations seem to be in agreement are the
$(S,M,T)=(0,0,1)$ and, perhaps, $(1,1,1)$ ones. In all the other channels there
are differences, more or less pronounced, in the shape of the responses, in the
position of the peak and in the behavior with respect to the HF results.  
The good agreement between the calculations at $q\simeq 0.1 k_F$, but not at
$q\simeq k_F$, seems to suggest that truncating the expansion of the residual
interaction at $l_\textrm{max}=3$ is not enough at large $q$. This conclusion
is independent on the presence of the tensor terms, since the disagreement
appears also in the $S=0$ channel and since the discrepancy survives also in
the comparison with the D1M parametrization (which does not contain tensor
terms), as seen from the results presented in the bottom right panel of our
Fig.~\ref{fig:resp_q270_kf270_d1snm} and in Fig.~5 of
Ref.~\cite{Pastore:2013sja}. 
One should mention that there are small differences in the choice of the
effective mass and of the momentum transfer between our figures and the ones of
Ref.~\cite{Pastore:2013sja}, as one can notice for example by the different end
point  ($\omega=\frac{q^2}{2m^{\star}}+\frac{qk_F}{m^{\star}}$) of the response
functions. We have repeated the calculations by choosing the same values of $q$
and $m^{\star}$ as Ref.~\cite{Pastore:2013sja}. The behaviour of our curves,
hence our conclusions, remain the same.

Turning to a global discussion of the responses obtained with the different
Gogny forces containing tensor terms, the following comments are in order:
\begin{itemize}
  \item{} One observes in general homogeneity of results in the $S=0$
    channels and some heterogeneity in the $S=1$ ones.

  \item{} The $(S,M,T)=(0,0,0)$ responses obtained with the different Gogny
    forces are in qualitative agreement among them, for $q=27$ $\textrm{MeV}/c$
    as well as for $q=270$ $\textrm{MeV}/c$, but differ from the ones obtained
    with the $G$-matrix interaction which were shown \cite{DePace:1996fb}  
    to successfully reproduce the $K^+$-nucleus quasielastic cross section,
    largely  dominated by the scalar-isoscalar channel, at $q=290, 390, 480$
    $\textrm{MeV}/c$.

  \item{} The $(S,M,T)=(0,0,1)$ responses as well present always
    similar features: they are characterized by a strong quenching of the
    continuum and the appearance of a collective mode at $q=27$
    $\textrm{MeV}/c$ and by some quenching and hardening (except for the GT2
    case) at $q=270$ $\textrm{MeV}/c$. 
    Similar features characterize also the $G$-matrix results, even if in this
    case at $q=27$ $\textrm{MeV}/c$ the collective mode enters in the
    continuum.

  \item{} Concerning the $S=1$ channels, a first remark is about the
    unphysical result of negative responses in some cases with the GT2 force,
    when calculated at first order in CF, which reflects a lack of convergence
    of the CF method when one considers this interaction. Furthermore, always
    for GT2, at the second order in the CF expansion all the $S=1$ responses
    are totally quenched and characterized by a collective mode at $q=27$
    $\textrm{MeV}/c$. This total quenching remains also at $q=270$
    $\textrm{MeV}/c$.

  \item{} For the other interactions one can observe that at $q=27$
    $\textrm{MeV}/c$ the differences in the $(S,M,T)=(1,1,0)$ and
    $(S,M,T)=(1,0,0)$ channels, for responses calculated with the D1ST* and
    D1MT* forces, essentially reflect the differences already present in the
    $(S,T)=(1,0)$ channel between D1S and D1M.   
    At $q=270$ $\textrm{MeV}/c$ the split between the $(S,M,T)=(1,1,0)$ and
    $(S,M,T)=(1,0,0)$ results is more or less pronounced depending on the
    interaction. 
    The two forces giving the results closer to the $G$-matrix calculations are, 
    for these channels, D1ST and D1MT. 

  \item{} The responses in $(S,M,T)=(1,0,1)$ channel always present at
    $q=27$ $\textrm{MeV}/c$ a collective mode above the continuum, also in the
    $G$-matrix case, while they can be very different from each other at $q=270$
    $\textrm{MeV}/c$.

  \item{}  In $(S,M,T)=(1,1,1)$ channel at $q=27$ $\textrm{MeV}/c$
    the collective mode can be above  or inside the continuum, depending on the
    interaction; at $q=270$ $\textrm{MeV}/c$, on the other hand, the qualitative
    behavior of all the responses is always very similar (except, as usual, for
    the GT2 case) and in agreement with the $G$-matrix results.
    This agreement opens the perspective of using these Gogny-type forces
    in the calculation of the neutrino-nucleus cross sections (and of the neutrino
    mean-free path in nuclear matter), which are dominated by the spin-isospin
    transverse response \cite{Martini:2009uj,Martini:2010ex,Pastore:2012dq}.  

 \item{} The differences between the results obtained at first and second order in the
CF expansion are in general more pronounced for these forces including tensor
terms with respect to the case of the standard parametrizations of the Gogny
forces.  
These differences are often very pronounced, in particular in the
$(S,M,T)=(1,0,1)$ case, for $q=27$ $\textrm{MeV}/c$ as well as for $q=270$
$\textrm{MeV}/c$ and remain also in the $\omega \to 0$ limit, important for the
calculations of the instabilities. 

\end{itemize}

\section{Finite size instabilities}
\label{sec:instabilities}

\begin{figure}
\begin{center}
\includegraphics[clip,width=0.48\textwidth]{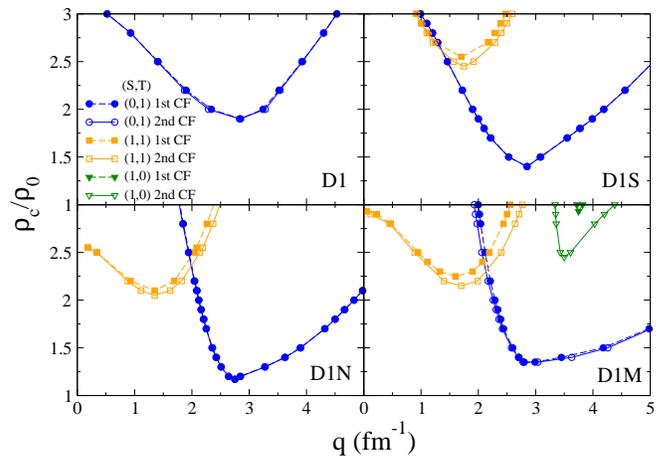}
\caption{(Color online) Critical densities $\rho_c$ divided by a constant value
  of the saturation density $\rho_0=0.16$ fm$^{-3}$ as a function of the
  transferred momentum $q$ (in fm$^{-1}$) for the most commonly used
  parametrizations of the Gogny force. The calculations of
  $R_{(S,T)}(q,\omega=0)$, through which the critical densities are deduced,
  are performed at first and second order in the continued fraction
  expansion. } 
\label{fig:rhoc_vsq_d1snm} 
\end{center}
\end{figure}

The study of the unphysical finite-size instabilities through the nuclear
matter response function formalism has attracted a lot of interest in the past
years starting from the work of Ref.~\cite{Lesinski:2006cu}.
Many recent investigations have been performed by considering Skyrme-type
nuclear energy density functionals
\cite{Lesinski:2006cu,Pastore:2012mf,Pastore:2012dq,Hellemans:2013bza,Pastore:2014yua,Pastore:2015vfa}.  
No studies of finite-size instabilities with the Gogny forces have been
published up to now. 

In the case of Skyrme functionals, in Ref.~\cite{Lesinski:2006cu} it has been
suggested a qualitative link between the appearance of finite-size
instabilities of nuclear matter near saturation density and the impossibility
to converge for self-consistent calculations in finite nuclei using some
parametrizations of the nuclear energy density functionals.
More precisely it has been shown that using
the SkP and LNS parameter sets the neutron and proton densities were
characterized by strong and opposing oscillating behavior, which increased with
the number of iterations of the self-consistent procedure.  
As the one-body equations of motion are solved iteratively, an instability in
the scalar isovector channel occurs when it becomes energetically favorable  
to build oscillations of neutrons against protons of unlimited amplitude. 
In Ref.~\cite{Lesinski:2006cu} it was also shown that the same SkP and LNS 
parametrizations lead to divergences of the nuclear matter response functions
at $\omega=0$ and finite $q$ when calculated in the $S=0$, $T=1$ channel for
densities close to the saturation one. These are the critical densities
$\rho_c$, \textit{i.e.} the lowest densities at which the nuclear response
calculated at zero transferred energy exhibits a pole.  

After the suggestion of the qualitative link between the finite nuclei and the
nuclear matter phenomena described above, several papers have been devoted to
the calculations of the critical densities at finite $q$ in nuclear and neutron
matter of many zero-range nuclear energy density functionals, including or not
the tensor components \cite{Pastore:2012mf,Pastore:2012dq,Pastore:2014yua}. 

A systematic quantitative analysis of the connection between the finite nuclei
and nuclear matter instabilities in the $S=0$, $T=1$ channel has been performed 
in Ref.~\cite{Hellemans:2013bza}, finding that a functional is stable if the
lowest critical density at which a pole occurs in nuclear matter calculations 
is larger than the central density of $^{40}$Ca, in practice around 1.2 times
the saturation density. In addition, one has also to verify that this pole
represents a distinct global minimum in the $(\rho_c,q)$ plane. This criterion
can be incorporated into the fitting procedure of the coupling constants of the
energy density functionals and has the advantage of being based on
computationally-friendly nuclear matter calculations.

In Ref.~\cite{Pastore:2015vfa} the quantitative analysis was extended to the
$S=1$ channel by studying not only ground-state properties, but also
vibrational excited states of finite nuclei.
The stability criterion mentioned above, derived in 
Ref.~\cite{Hellemans:2013bza} for the $S=0$, $T=1$, was found to remain valid
also in the $S=1$ channel. We remind that the $S=0$, $T=0$ channel is
characterized by the physical spinodal instability, hence it is not considered in
the studies of unphysical instabilities.

All the studies described above considered Skyrme functionals.
Turning to the Gogny interaction, the stability studies involving nuclear
matter calculations have considered up to now only the $q \to 0$ limit,
corresponding to perturbation of infinite wavelength, and have used the
well-known stability conditions established by Migdal \cite{Migdal_book},
starting from the seminal work of Gogny and Padjen \cite{GognyNPA77}.
In this context, beyond the $(S,T)=(0,0)$ channel, where the stability
condition is not satisfied at low densities, reflecting the existence of the
well known spinodal instability, in the other channels the infinite wavelength
($q=0$) instabilities of the most commonly used Gogny interactions typically
appears for densities larger than two or three times the nuclear matter
saturation density $\rho_0$. For example in the D1N case the lowest density
instability appears in the $(1,1)$ channel at $\rho_C \simeq 2.5 \rho_0$, while
for D1S it appears in the $(0,1)$ channel at $\rho_C \simeq 3.5 \rho_0$.

Here we consider for the first time the evolution of the critical densities
with the momentum transfer $q$. 
Discarding the $(0,0)$ channel and its corresponding spinodal
instability, we start by showing in Fig.~\ref{fig:rhoc_vsq_d1snm} the critical
densities in the other $(S,T)$ channels as a function of $q$ for the four most
commonly used parametrizations of the Gogny force.

In the D1 case only the $R_{(0,1)}(q,\omega=0)$ exhibits a pole at finite $q$.
The corresponding critical density is never lower than $2 \rho_0$.
For the other three Gogny parametrizations (D1S, D1N and D1M) the
poles appears at finite $q$ not only in the $(0,1)$ channel, but also in the
$(1,1)$ one (and in the $(1,0)$ for D1M). This $(1,1)$ channel, even if it
presents critical densities lower that $3 
\rho_0$ already for $q=0$ in the case of D1N and D1M, is characterized by a
relatively smooth decrease of $\rho_C$ with $q$ and in any case for the D1S,
D1N and D1M the corresponding critical densities are never lower than $2
\rho_0$, even at large $q$. On the contrary, for the $(0,1)$ channel the
critical densities rapidly decrease with $q$ reaching values around
$\rho_C\simeq1.5 \rho_0$ for D1S and D1M and around $\rho_C\simeq1.2 \rho_0$
for D1N. All the curves obtained are furthermore characterized by a global
minimum in the $(\rho_c,q)$ plane. 
The two stability criteria established in Ref.~\cite{Hellemans:2013bza} are
thus satisfied by the most commonly used Gogny forces. Hence they are free
from instability problems. The only case to be treated with some caution
is the one of the D1N parametrization, which in any case is rarely used in the
finite nuclei calculations. The case of D1N, together with D1M, presents also
some small differences in the critical densities  between the results obtained
by calculating the nuclear response functions at the first order in the CF
expansion and the results at second one, while for the D1 and D1S case the
results at the two orders practically coincide. For all these four interactions
we can anyway conclude that the calculation at the first order in the CF
expansion can be considered enough for finite-size instabilities studies.   

\begin{widetext}

\begin{figure}
\begin{center}
\includegraphics[clip,width=\textwidth]{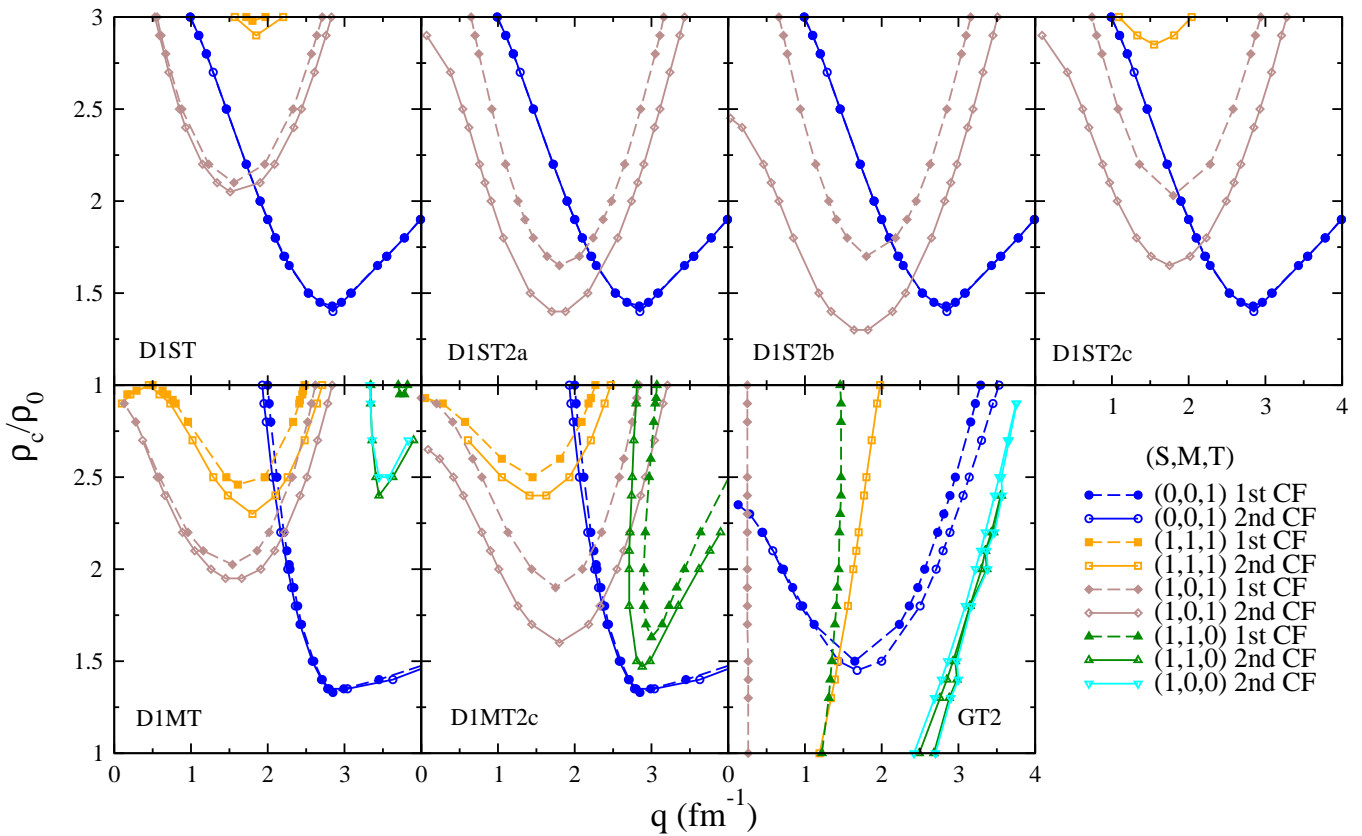}
\caption{(Color online) As in Fig.~\ref{fig:rhoc_vsq_d1snm}, but for the Gogny
  forces with tensor terms.}
\label{fig:rhoc_vsq_gognytens} 
\end{center}
\end{figure}

\end{widetext}

Turning to the parametrizations of the Gogny forces including tensor terms, we
show in Fig. \ref{fig:rhoc_vsq_gognytens} the results for the critical
densities as a function of the momentum transfer calculated at first and second
order in the CF expansions for the D1ST, D1ST2(a,b,c) D1MT, D1MT2c and GT2
interactions. 

In the case of D1MT, in the $q \to 0$ limit  we find for the critical density in
the (1,1) channel a value close to $\rho_c=0.45$ fm$^{-3}$,  a result already
obtained in Ref.~\cite{Navarro:2013bda} in the context of spin susceptibilities
calculations. This is the only point with which we can compare our calculations
including tensor terms.

As in the case without tensor terms, finite size instabilities appear only in
the $T=1$ channels for the D1ST*-type force, while they appear in the $T=1$
channels and also for $(S,M,T)=(1,1,0)$ for the D1MT*-type forces. The behavior
of the critical densities in the $(S,M,T)=(0,0,1)$ channel for the  D1ST*- and
D1MT*-type forces reflects the one of D1S and D1M. In the $S=1$, $T=1$ channel
the presence of tensor terms lower (raise) the values of the critical densities
for the $M=0$ ($M=1$) component, when compared to the corresponding cases
without tensor.

In the $S=1$ channels differences between the results obtained at the first and
second orders in the CF expansion appear. 
In any case  the two stability criteria established in
Ref.~\cite{Hellemans:2013bza} are always satisfied by all the parametrization
of the D1ST*- and D1MT*-type Gogny forces considered here in all the  $(S,M,T)$
channels. 

This reassuring results no longer hold for the GT2 force  
which presents finite $q$ instabilities at any density in the $(S,M,T)=(1,1,0)$
and $(S,M,T)=(1,0,1)$ channels.  
These results are the counterpart of the peculiar behavior of the response
functions calculated with this GT2 force: negative at first order in the CF
expansion and with a ph continuum totally quenched at second order in the CF
expansion.

\section{Conclusion}

We have studied RPA nuclear matter response functions by considering the
nucleons interacting via the finite range Gogny force.  
We have considered the most commonly used parametrizations of this force, as
well as some recent generalizations that include the tensor terms.
We have performed a fully antisymmetrized RPA calculation, including
\textit{i.e.} the exchange contribution of the particle-hole interaction,  
by employing the continued fraction (CF) technique. The calculations have been
performed by truncating the continued fraction expansion at first and second
order, the highest one so far reached in the context of finite range forces. 

Concerning the most commonly used parameter sets of the Gogny force, we have
obtained results similar to the ones presented in Ref.~\cite{Margueron:2008dq}  
for the D1 interaction, the only interaction considered in that paper. 
The response functions in the four $(S,T)$ channels obtained with the other
common parameter sets, namely D1S, D1N and D1M, have been presented here for
the first time. 
Some differences in the transferred-energy behavior appear, in particular at
small momentum transfer, depending on the chosen parametrization.  
The convergence of the CF expansion as well strongly depends on the force
parameters.

Concerning the Gogny forces including tensor terms, we have considered the
D1ST, D1ST2(a,b,c) D1MT, D1MT2c and GT2 interactions. 
In the case of the D1MT interaction we have compared our results with the
ones published in Ref.~\cite{Pastore:2013sja}, where the responses for the D1MT
interaction have been calculated in the Landau framework by truncating the
residual interaction up to $l_\textrm{max}=3$. For all the other interactions
including the tensor terms, the response functions appear here for the first
time. 
A general homogeneity characterizes the $S=0$ channels, while in the $S=1$ an
heterogeneity of the results appears, reflecting the very different choices
for the tensor terms.  Furthermore, the differences between the results obtained
at first and second order in the CF expansion turn to be more pronounced for
the forces including tensor terms with respect to the case of the standard
Gogny ones.

An interesting point is the behavior of the responses calculated with
the difference forces including the tensor terms in the $(S,M,T)=(1,1,1)$
channel,  which is similar for all the parametrizations and in 
agreement with the $G$-matrix results. This agreement opens the
perspective of using these Gogny-type forces in the calculation of the
neutrino-nucleus cross sections (and of the neutrino mean-free path in nuclear
matter) which are dominated by the spin-isospin transverse response. 
The correspondence between continuum RPA calculations in finite nuclei
and nuclear matter results in the quasielastic electron and neutrino scattering
has been illustrated for example in Refs.~\cite{Bauer:2000wz,Martini:2016eec}. 

Continuum RPA calculations in finite nuclei using the Gogny force have
been developed in Ref.~\cite{DeDonno:2010za} and recently generalized to the
charge-exchange excitations in Ref.~\cite{DeDonno:2016vge}.  
It would be very interesting to compare the finding of that approach with the
one developed in the present paper (employing a same Gogny interaction with or
without tensor terms) in order to investigate up to where one can push a
nuclear matter approach to study finite nuclei properties and reactions. 

As a first instance of connection between nuclear matter and finite
nuclei, we have employed the present nuclear matter approach to study the
spurious finite-size instabilities. 
We have shown for the first time that, at variance with some zero
range Skyrme functionals, the most commonly used finite range D1, D1S, D1M
Gogny forces satisfy the stability criteria of Ref.~\cite{Hellemans:2013bza} in all the $(S,T)$ channels, 
hence they are free of spurious finite-size instabilities. 
The only case to be treated with some caution
is the one of the D1N parametrization, which in any case is rarely used in the
finite nuclei calculations. 

The stability criteria of Ref.~\cite{Hellemans:2013bza} 
are also satisfied in all the $(S,M,T)$ channels by all the Gogny forces including tensor terms of type
D1MT* and D1ST*. On the contrary, at least at first and second order in the CF expansion, the
GT2 force is unstable in all the $S=1$ channels. 
However, this results might just signal the poor convergence of the CF
  expansion for the GT2 interaction, given the extremely strong tensor
  component which is present in this force. 

In perspective, the tool developed in the present paper could be
inserted in the fitting procedure to construct new Gogny-type forces.  
It would be also interesting to repeat the study of the present paper by
considering the recently developed D2 Gogny force \cite{Chappert:2015eaa},
characterized by a finite range density dependent term, as well as other finite
range forces \cite{Bertsch:1977sg,Nakada:2003fw,Dobaczewski:2012cv} employed
for low-energy nuclear structure calculations.

\appendix*
\section*{Appendix: Continued fractions with the Gogny interaction}

We give here the explicit expressions for the first and second order exchange 
diagrams, based on the potential of Eq. (\ref{eq:pot}), where for $V_\alpha(k)$ we
take the gaussian expression typical of the Gogny forces, the label $\alpha$
standing for the spin-isospin channel.

After some manipulations, as explained in detail in Ref.~\cite{DePace:1998yx},
the first order polarization propagator can be cast into the following form:
\begin{eqnarray}
  \Pi^{(1)\text{ex}}_{\alpha}(q,\omega) &=&
    -\left(\frac{m}{q}\right)^{2}\frac{k_F^4}{(2\pi)^4}
    \left[{\cal Q}_\alpha^{(1)}(0,\psi)
    - {\cal Q}_\alpha^{(1)}(\bar{q},\psi) \right. \nonumber \\
  &&  \left. + {\cal Q}_\alpha^{(1)}(0,\psi+\bar{q})
    - {\cal Q}_\alpha^{(1)}(-\bar{q},\psi+\bar{q})\right] , \nonumber \\
\end{eqnarray}
where 
\begin{eqnarray}
  {\cal Q}_\alpha^{(1)}(\bar{q},\psi) &=& 2 \int_{-1}^1 dy
    \frac{1}{\psi-y+i\eta_{\omega}} \\
  && \times \int_{-1}^1 dy' \, {W_\alpha}''(y,y';\bar{q})
    \frac{1}{y-y'+\bar{q}} , \nonumber 
\end{eqnarray}
whereas for the second order polarization propagator one has:
\begin{widetext}
\begin{eqnarray}
  && \Pi^{(2)\text{ex}}_{\alpha\alpha'}(q,\omega) =
    \left(\frac{m}{q}\right)^{3}\frac{k_F^6}{(2\pi)^6}
    \left[{\cal Q}_{\alpha\alpha'}^{(2)}(0,0;\psi)
        - {\cal Q}_{\alpha\alpha'}^{(2)}(0,\bar{q};\psi)
        - {\cal Q}_{\alpha\alpha'}^{(2)}(\bar{q},0;\psi)
        + {\cal Q}_{\alpha\alpha'}^{(2)}(\bar{q},\bar{q};\psi) \right.
  \nonumber \\
  &&\qquad \left.
        - {\cal Q}_{\alpha\alpha'}^{(2)}(0,0;\psi+\bar{q})
        + {\cal Q}_{\alpha\alpha'}^{(2)}(0,-\bar{q};\psi+\bar{q})
        + {\cal Q}_{\alpha\alpha'}^{(2)}(-\bar{q},0;\psi+\bar{q})
        - {\cal Q}_{\alpha\alpha'}^{(2)}(-\bar{q},-\bar{q};\psi+\bar{q})
    \right] , \nonumber \\
\end{eqnarray}
\end{widetext}
where \footnote{This expression is valid for central-central and
  central-tensor terms. About tensor-tensor contributions see below.}
\begin{eqnarray}
  {\cal Q}_{\alpha\alpha'}^{(2)}(\bar{q}_1,\bar{q}_2;\psi) &=& \int_{-1}^1 dy
    \frac{1}{2}\int_{0}^{1-y^2} \!\!\! dx 
    {\cal G}_{\alpha}(x,y+\bar{q}_1;\psi+\bar{q}_1) \nonumber \\
  && \times \frac{1}{\psi-y+i\eta_{\omega}}
    {\cal G}_{\alpha'}(x,y+\bar{q}_2;\psi+\bar{q}_2) \nonumber \\
  \label{Eq:Q2}
\end{eqnarray}
and
\begin{equation}
  {\cal G}_{\alpha}(x,y;\psi) = \int_{-1}^1 dy'
    \frac{1}{\psi-y'+i\eta_{\omega}}W'_\alpha(x,y;y') .
\end{equation}
In the above expressions $\eta_\omega\equiv\text{sign}(\omega)\eta$,
$\bar{q}\equiv q/k_F$ is the transferred momentum in units of $k_F$ and $\psi$
the Fermi gas scaling variable, which reads, for non-relativistic kinematics,
\begin{equation}
  \psi = \frac{1}{k_F}\left(\frac{\omega m}{q}-\frac{q}{2}\right).
\end{equation}
The auxiliary functions $W'$ and $W''$ are given in terms of the integral over
the azimuthal angle of the momentum dependent part of the interaction as
\begin{eqnarray}
  W'_\alpha(x,y;y') &=& \frac{1}{2}\int_0^{1-{y'}^2}\!\!\!\!\! dx' 
  W_\alpha(x,y;x',y')
  \label{Eq:Wpalpha} \\
  W''_\alpha(y,y';\bar{q}) &=& \frac{1}{2}\int_0^{1-y^2}\!\!\!\!\! dx\,
    \frac{1}{2}\int_0^{1-{y'}^2}\!\!\!\!\!\! dx' W_\alpha(x,y+\bar{q};x',y') , 
    \nonumber \\ \label{Eq:Wppalpha}
\end{eqnarray}
where 
\begin{eqnarray}
  W_{\alpha}(x,y;x',y') &=& \int_0^{2\pi}\frac{d\varphi}{2\pi}
  V_{\alpha}(\bm{k}-\bm{k}')\phantom{S_{zz}(\widehat{\bm{k}-\bm{k}'})}
  \label{Eq:Walpha}
    \\
  W_{\alpha}(x,y;x',y') &=& \int_0^{2\pi}\frac{d\varphi}{2\pi}
    V_{\alpha}(\bm{k}-\bm{k}') S_{zz}(\widehat{\bm{k}-\bm{k}'}) ,
    \nonumber \\ \label{Eq:Walphatn}
\end{eqnarray}
for the non-tensor and tensor terms, respectively, and 
$S_{zz}(\hat{k})=3\hat{\bm{k}}_z \hat{\bm{k}}_z-1$.
The $x$ and $y$ variables are defined in terms of the momentum ${\bf k}$ as
$y=k\cos\theta$ and $x=k^2-y^2$.

Employing a gaussian interaction, only Eq.~(\ref{Eq:Walpha}) can be calculated
analytically (in terms of a modified Bessel function of first kind), so that in
general one has to cope with multidimensional numerical integrations in
presence of (integrable) singularities.
We have calculated these integrals using a mix of deterministic and Monte Carlo
\cite{Cuba05} techniques: reaching a good accuracy turns out to be quite time
consuming. A faster way of performing these calculations is provided by fitting
the Gogny potentials in terms of meson exchanges.
It turns out that a good fit of all the potentials employed in this work, for
momenta up to 1~GeV/c, can be obtained through the exchange of the $\pi$,
$\sigma$, $\rho$ and $\omega$ mesons, with standard dipole form factors, using
the four coupling constants as fitting parameters.
By using meson-exchange potentials,
Eqs.~(\ref{Eq:Wpalpha})--(\ref{Eq:Walphatn}) can be calculated
analytically (see Ref.~\cite{DePace:1998yx} for details), with a substantial
improvement in computing time.
As a cross-check the calculations shown in the paper have been performed using
both techniques.

Here, for completeness, we provide also the explicit expressions for the
second order tensor-tensor contributions in the CF expansion, which had not
been shown in Ref.~\cite{DePace:1998yx}.
When both the interaction lines in Fig.~\ref{fig:RPA}c contain a tensor term,
the resulting contributions cannot be expressed in a factorized form as in
Eq.~(\ref{Eq:Q2}), but rather as a sum of factorized terms:
\begin{eqnarray}
  && {\cal Q}_{\alpha\alpha'}^{(2)}(\bar{q}_1,\bar{q}_2;\psi) = 2 \int_{-1}^1 dy
    \frac{1}{2}\int_{0}^{1-y^2} \!\!\! dx \frac{1}{\psi-y+i\eta_{\omega}} 
    \nonumber \\
  && \quad\times \left[ {\cal G}^{(a)}_{\alpha}(x,y+\bar{q}_1;\psi+\bar{q}_1)
      {\cal G}^{(a)}_{\alpha'}(x,y+\bar{q}_2;\psi+\bar{q}_2) \right.\nonumber \\
  && \qquad + {\cal G}^{(b)}_{\alpha}(x,y+\bar{q}_1;\psi+\bar{q}_1)
      {\cal G}^{(c)}_{\alpha'}(x,y+\bar{q}_2;\psi+\bar{q}_2) \nonumber \\
  && \qquad + {\cal G}^{(c)}_{\alpha}(x,y+\bar{q}_1;\psi+\bar{q}_1)
      {\cal G}^{(b)}_{\alpha'}(x,y+\bar{q}_2;\psi+\bar{q}_2) \nonumber \\
  && \qquad - \left. {\cal G}^{(d)}_{\alpha}(x,y+\bar{q}_1;\psi+\bar{q}_1)
      {\cal G}^{(d)}_{\alpha'}(x,y+\bar{q}_2;\psi+\bar{q}_2) \right]\nonumber \\
  && \quad \alpha,\alpha'=t,t\tau \label{Eq:Q2tens}
\end{eqnarray}
with
\begin{eqnarray}
  {\cal G}^{(l)}_{\alpha}(x,y;\psi) &=& \int_{-1}^1 dy'
    \frac{1}{\psi-y'+i\eta_{\omega}}W'^{(l)}_\alpha(x,y;y') , \nonumber \\ \\
    W'^{(l)}_\alpha(x,y;y') &=& \frac{1}{2}\int_0^{1-{y'}^2}\!\!\!\!\! dx' 
  W^{(l)}_\alpha(x,y;x',y') \label{Eq:Wptens}
\end{eqnarray}
and
\begin{eqnarray}
  W^{(a)}_\alpha(x,y;x',y') &=& \int_0^{2\pi}\frac{d\varphi}{2\pi}
    V_{\alpha}(\bm{k}-\bm{k}')
    \frac{h()-2(y'-y)^2}{h()+(y'-y)^2} \nonumber \\ \\
  W^{(b)}_\alpha(x,y;x',y') &=& \int_0^{2\pi}\frac{d\varphi}{2\pi}
    V_{\alpha}(\bm{k}-\bm{k}') \\
  W^{(c)}_\alpha(x,y;x',y') &=& \int_0^{2\pi}\frac{d\varphi}{2\pi}
    V_{\alpha}(\bm{k}-\bm{k}')
    \frac{2 h()-(y'-y)^2}{h()+(y'-y)^2} \nonumber \\ \\
  W^{(d)}_\alpha(x,y;x',y') &=& \int_0^{2\pi}\frac{d\varphi}{2\pi}
    V_{\alpha}(\bm{k}-\bm{k}')
    \frac{h() (y'-y)}{h()+(y'-y)^2}, \nonumber \\ \label{Eq:Wd} 
\end{eqnarray}
having defined
\begin{equation}
  h() = h(x,x',\varphi) = x'+x-2\sqrt{x'}\sqrt{x}\cos(\varphi).
\end{equation}
Again, for gaussian potentials all the integrations have to be performed
numerically, whereas for meson-exchange potentials the expressions
(\ref{Eq:Wptens})--(\ref{Eq:Wd}) can be obtained analytically.

\acknowledgments
M.M. acknowledges the support and the framework of the ``Espace de Structure et
de r\'eactions Nucl\'eaire Th\'eorique'' (ESNT, \url{http://esnt.cea.fr} ) at
CEA as well as the hospitality of the INFN-Torino where part of this work was
done.

\end{document}